\documentstyle[twoside,fleqn,espcrc2]{article}

\newcommand{\AmS}{{\protect\the\textfont2
  A\kern-.1667em\lower.5ex\hbox{M}\kern-.125emS}}
\title{Hamiltonian lattice gauge theory: wavefunctions on large lattices}

\author{J. B. Bronzan\address{Department of Physics and Astronomy,
        Rutgers University \\
        P.O. Box 849, Piscataway, New Jersey 08855-0849, U.S.A.}%
        \thanks{Work supported in part by the U.S. National Science
        Foundation under grant number PHY 88-18535.}}

\begin{document}

\begin{abstract}
We discuss an algorithm for the approximate solution of Schr\"{o}dinger's
equation for lattice gauge theory, using lattice SU(3) as an example.  A
basis is generated by repeatedly applying an effective Hamiltonian to a
``starting state.''  The resulting basis has a cluster decomposition and
long-range correlations.  One such basis has about \(10^4\) states on a \(10
\times 10\times 10\) lattice.  The Hamiltonian matrix on the basis is sparse,
and the elements can be calculated rapidly.  The lowest eigenstates of the
system are readily calculable.

\end{abstract}

\maketitle

The approximate solution of Schr\"{o}dinger's equation for lattice gauge
theory presents a number of problems, especially when it is to be carried out
on a large lattice.  We discuss these issues for pure SU(3) field theory,
sometimes using the example of a \(10\times 10\times 10\) lattice, where there
are 3000 link degrees of freedom.  The addition of matter degrees of freedom
is straightforward.

\section{CHOICE OF THE HAMILTONIAN}

The computations of SU(3) wavefunctions is based on the Kogut-Susskind
Hamiltonian for SU(3) lattice gauge theory.\cite{Kogut}
\begin{eqnarray}
\lefteqn{H=\frac{g^2}{2}\sum_{{\bf s},\mu}{\cal J}^2_{{\bf s},\mu}-
\frac{1}{g^2}\sum_{{\bf s},\mu >\nu}} \label{eq:ks} \\
&&\times\left[ tr\left(u_{{\bf s},\mu}
u_{{\bf s}+\hat\mu,\nu}u^{\dagger}_{{\bf s}+\hat\nu,\mu}u^{\dagger}_{{\bf s},
\nu}\right)+H.c.-6\right].\nonumber
\end{eqnarray}
This Hamiltonian commutes with local gauge transfromations, and it is only its
locally gauge-invariant eigenstates that are physically significant.  One
way to eliminate states having color charge on the lattice is to fix the
gauge.  The resulting Hamiltonian has fewer degrees of freedom, but is
very messy.\cite{Gafix}  For example, the electric term in the
Hamiltonian becomes
\begin{eqnarray}
\lefteqn{H_E=\frac{g^2}{2}\sum_{\alpha ,a,b,c}\{ \beta_1
{\cal J}_{L,\alpha }(y_{a1,b1,c1}){\cal J}_{L,\alpha }(y_{a2,b2,c2})}
\nonumber \\
&&+\beta_2{\cal J}_{L,\alpha }(z_{a1,b1,c1}){\cal J}_{L,\alpha }(z_{a2,b2,c2})
\nonumber \\
&&+\beta_3{\cal J}_{R,\alpha }(z_{a1,b1,c1}){\cal J}_{R,\alpha }(z_{a2,b2,c2})
\nonumber \\
&&+\beta_4\left[{\cal J}_{L,\alpha }(y_{a1,b1,c1})+{\cal J}_{L,\alpha }
(z_{a1,b1,c1})\right]\nonumber \\
&&\times\left[{\cal J}_{L,\alpha }(y_{a2,b2,c2})+
{\cal J}_{L,\alpha }(z_{a2,b2,c2})\right]\nonumber \\
&&+\beta_5\left[{\cal J}_{R,\alpha }(y_{a1,b1,c1})+{\cal J}_{R,\alpha }
(z_{a1,b1,c1})\right]\nonumber \\
&&\times\left[{\cal J}_{R,\alpha }(y_{a2,b2,c2})+
{\cal J}_{R,\alpha }(z_{a2,b2,c2})\right]\nonumber \\
&&+\beta_6\left[{\cal J}_{R,\alpha }(z_{a1,b1,c1})-
{\cal J}_{L,\alpha }(z_{a1,b1,c1})\right]\nonumber \\
&&\times {\cal J}_{L,\alpha }(z_{a2,b2,c2})\}.
\label{eq:fix}
\end{eqnarray}
The functions $\beta$ are nonzero for degrees of freedom located arbitrarily
far from one another.  Here is one of them:
\begin{equation}
\beta_4(a_1,b_1,c_1;a_2,b_2,c_2)=min(c1,c2).
\end{equation}

The locality and homogeneity of the Kogut Susskind Hamiltonian Eq.~\ref{eq:ks}
have been lost in Eq.~\ref{eq:fix}.  However, these characteristics are
crucial to our construction of basis states, and we therefore opt to use
Eq.~\ref{eq:ks}.  This implies that the states we construct will be
contaminated with color.  Our strategy is to construct
eigenstates of Eq.~\ref{eq:ks} for a range of expectation values
of the lattice color operator \(\hat Q^2\equiv\sum_i \hat Q_i^2\).  (The
sum is over sites \(i\), where \(\hat Q_{i,\alpha}\) are the generators of
gauge transformations.)  We then extrapolate our results to color zero.

\section{BASIS STATES}

Basis states are products of factors \(\phi\) or \(\phi_{\alpha}\) for
each degree of freedom.  These factors are ``harmonic oscillator'' states
on the SU(3) manifold.\cite{Harm1}  They transform as scalars
and octets, respectively, under SU(3) transformations.  Higher tensor
states exist, but have proven unimportant in work on small lattices.
We omit them at this stage of our work.

The construction of a basis begins with a ``starting state'' \(|0\rangle\)
in which each link has the same factor \(\phi\).  This state has a parameter
in, the width of the harmonic oscillator Gaussian, and it is chosen to
minimize the expectation of \(H\) at the coupling we choose.  At strong
coupling the state is the true ground state.  At weak coupling the
expectation value is 26\% above the ground state energy.  At weak and
intermediate couplings the starting state has nonzero color density,
which must be eliminated later by extrapolation.

It is not feasible to add basis states by using the nine states \(\phi,
\phi_{\alpha}\) as a basis on each link.  On a \(10\times 10\times 10\)
lattice there are 3000 links and therefore \(9^{3000}\) lattice states are
generated in this manner.  It is better to set up a basis using dynamical
information by making the following query:  If \(|\Phi\rangle\) is a lattice
state, what is the optimal state \(\hat A|\Phi\rangle\) to add so that a
linear combination can be formed having given color \(\langle\hat Q^2
\rangle\) and minimum energy?  The (approximate) variational choice for
\(\hat A\) is
\begin{equation}
\hat A=H+\lambda\hat Q^2.
\end{equation}
Here \(\lambda\) is a parameter.  As we tune it, we change both \(\langle\hat
 Q^2\rangle\) and the variational energy.  The rest of the basis is generated
by repeated application of \(\hat A\) to the starting state:
\begin{equation}
|\Phi_n\rangle =(\hat A)^n|0\rangle .  \label{eq:basis}
\end{equation}

\section{EXTRAPOLATION TO ZERO COLOR}

Using basis~\ref{eq:basis}, it is possible to make a simple test of the
extrapolation procedure to zero color.  We do this at weak coupling where
perturbation theory provides a comparison benchmark.  The basis we use
is the smallest one that allows study of the problem: \(\Phi_0\) and
\(\Phi_1\).  It is convenient to write the color and energy in terms of
variables \(x\) and \(y\).
\begin{equation}
\langle\hat Q^2\rangle=\frac{4L^3}{g^2}+\frac{16L^{3/2}}{g^2}x,
\end{equation}
\[E=24L^3+3L^{3/2}y.\]
\(L\) is the number of links on the edge of the lattice.  Variables \(x,y\)
fall on a tilted eccentric ellipse (\(b/a=0.41\)), given by the equation
\begin{equation}
8x^2-16xy+21y^2=13.
\end{equation}
As \(\lambda\) sweeps from \(-\infty\) to \(+\infty\), \(x\) varies from
\(+\sqrt{21/8}\) to \(-\sqrt{21/8}\), and \(y\) moves over the lower branch
of the ellipse from \(+\sqrt{8/21}\) to \(-\sqrt{8/21}\).  The simplest
extrapolation is to use a straight line passing through these endpoints
to reach \(\langle\hat Q^2\rangle =0\).  When this is done, the extrapolated
ground state energy extrapolated to the physical sector becomes \(E_0=144L^3/
7\), which is 8\% above the exact weak-coupling result \(E_0=19.1L^3\).
This is to be compared with the 26\% excess when only starting state
\(\Phi_0\) is present.  (This state has color density \(\langle\hat Q^2\rangle
/L^3=48/g^2\).)

\section{THE CLUSTER EXPANSION}

The Kogut-Susskind Hamiltonian is a sum of operators that are either local or
nearly local, and the same is true of the color operator \(\hat Q^2\).  We
denote these terms generically as \(A_i\); then Eq.~\ref{eq:basis} becomes
\begin{equation}
|\Phi_n\rangle =\sum_{i1,\dots ,in}A_{i1}A_{i2}\cdots A_{in}|0\rangle .
\end{equation}
Terms involving different links commute, so we can rearrange this into the
form
\begin{equation}
|\Phi_n\rangle =\sum_{i1,\dots ,in}[A_{ia}A_{ib}\cdots ][A_{i\alpha}
A_{i\beta}\cdots ]\cdots |0\rangle ,  \label{eq:cluster}
\end{equation}
where the indices are a permutation of \(i1,\dots ,in\), and the factors
within a given bracket share a degree of freedom with another factor
within the same bracket, but not with any factor in other brackets.  These
bracketed factors produce clusters of excitation of the harmonic oscillator
state \(\phi\to\phi_{\alpha}\) on the lattice.  {\it The cluster
representation is a useful framework for constructing a basis because A is
a sum of nearly local and translation invariant operators.  Clusters are
therefore small (locality) and identical wherever they appear on the
lattice (translation invariance).}  Our approach is to compute and store
cluster amplitudes, and then use an elaboration of Eq.~\ref{eq:cluster} to
compute lattice wavefunctions from them.  This approach is not available with
the gauge fixed Hamiltonian~\ref{eq:fix}.

There is one further deomposition of \(\Phi_n\) that we choose to make.
In Eq.~\ref{eq:cluster} different index assignments lead to different
numbers of clusters.  We separately group the terms having \(n_c\) clusters
into a substate \(\Phi_{n,n_c}\) and retain these different linear
combinations as separate cluster states.

When a single term in operator \(A\) is applied to the starting state, it
can have several effects, including the creation of a state with a pair
of excitations having contracted color indices.
\begin{equation}
A_i|0\rangle =\sum a\phi(1)\cdots\phi_{\alpha}(j)\cdots\phi_{\alpha}(k)
\cdots ,
\end{equation}
where \(j\) and \(k\) designate links appearing in \(A_i\).  Application
of a second term in \(\hat A\) can either change the first pair, or
create a second cluster.  A particularly important change occurs when the
second application extends the distance between links \(j\) and \(k\) to
build longer-range correlations.  This is a contribution to \(\Phi_{2,1}\)
and we know that such terms are crucial in the scaling region.  Unfortunately,
such long-range terms are apt to be swamped in the matrix elements.  For
example, in the norm of the state \(\Phi_2\), the contribution of
\(\Phi_{2,2}\) dominates the contribution of \(\Phi_{2,1}\) by a factor of
the lattice volume because the two clusters can range independently over
the lattice.

We therefore make adjustments to the basis implied by Eq.~\ref{eq:basis}:

1) As mentioned, we separate \(\Phi_n\) into distinct states \(\Phi_{n,n_c}\)
to evade swamping.

2) In each cluster, we retain only a one-pair excitation like that
discussed.  When this is done, clusters are the same thing as pairs.  Pairs
are computed iteratively.  That is, to obtain the pairs of order \(N\), we
apply \(\hat A\) to the clusters of order \(N-1\), and store amplitudes.
The number of distinct kinds of pairs of clusters can be no more than
\(3L(L+1)^2/2\); this number of clusters presents no computational problem
on a \(10\times 10\times 10\) lattice.

The basis we construct in this manner is a stripped-down version of Eq.~\ref
{eq:basis}, but it still is constructed dynamically, and contains long-range
correlations.  We believe these features are the essential ones for correct
scaling.  In addition, we believe it is important that our starting state is
reasonably accurate at weak and strong coupling.

We emphasize that the lattice states we construct have many pair excitations.
For example, the state \(\Phi_{137,111}\) has 111 pairs on the lattice.  A
related point is that when \(n_c=3L(L+1)^2/2\), {\it every} link on the
lattice is excited.  We include such states in our basis because the only
way we can shift the ground state energy density is to tile the lattice
with excitations.

We anticipate including cluster states having \(n\le 3L(L+1)^2/2\) and
\(n-n_c=N\le L\).  The basis therefore grows like \(L^4\).

\section{COMPUTABILITY OF MATRIX ELEMENTS}

The lattice state \(\Phi_{n,n_c}\) is a sum over states in which \(n_c\)
pairs are placed on the lattice in various ways.  It is not immediately
apparent that it is feasible to compute matrix elements of these states
because of the enormous number of terms to be considered.  Two features
are of help here, and they can be discussed using the case of the computation
of state norms.

1) We saw in Section 4 that phase space favors
those terms in which a maximum number of pairs can be positioned
independently on the lattice.  In the case of the norm, contributions in
which the set of pairs in the bra is identical to the set of pairs in
the ket dominate other contributions by at least one power of the lattice
volume.  We retain only these leading contributions.

2) There is still danger that the norm will be uncomputable.  It has the
stucture
\begin{equation}
\langle\Phi_{n,n_c}|\Phi_{n,n_c}\rangle =\sum_{pairs}a(\{i\})
\Big |_{\sum_i N_i=n}.
\label{eq:norm}
\end{equation}
Here \(i\) enumerates pair type, \(N_i\) is the number of applications
of \(\hat A\) required to produce pair \(i\), and there are \(n_c\) pair
types in each contribution.  \(a(\{i\})\) is an amplitude assembled from the
pair amplitudes \(x_i\) computed by the iterative procedure outlined in
Section 4.  The number of pair types is of order
\(L^3\), so the number of terms in Eq.~\ref{eq:norm} is the same as the
number of partitions of n into \(L^3\) non-negative integers, or
\begin{equation}
\frac{(n+L^3)!}{(L^3-1)!n!}.\nonumber
\end{equation}
This is an impossibly large number on a lattice of any size.  Fortunately,
the amplitudes \(a\) in Eq.~\ref{eq:norm} depend on the pair amplitudes in
a manner that allows us to use the distributive laws of arithmetic to sum
before multiplying the pair amplitudes.  This possibility is illustrated
by the choice
\begin{equation}
a=\prod_i(x_i)^{N_i}/(N_i)!
\end{equation}
Then
\begin{equation}
\langle\Phi_n|\Phi_n\rangle =\frac{1}{n!}\left (\sum_i x_i\right )^n.
\end{equation}
Similar simplifications occur in the computation of other matrix elements of
\(\hat A\).  These simplifications are discovered by writing generating
functions for matrix elements.  Manipulation of these matrix elements
teach us how to interchange multiplication
and summation of pair amplitudes.  Preliminary numerical work on a
\(10\times 10\times 10\) lattice indicates that the computation of the
Hamiltonian matrix is not taxing.

\section{CONCLUSIONS}

We are working at present on a \(10\times 10\times 10\) lattice, in which
\(n_c\) ranges up to 1,500, so that the last states tile the entire lattice
with pairs.  We allow \(n\) to exceed \(n_c\) by as much as 6, so that
pairs separated by up to 6 lattice spacings appear on the lattice.  There are
thus 9001 states in our basis.  The Hamiltonian has \(\sim4\times 10^7\)
distinct matrix elements, but of these only 162,000 are nonzero.
On a workstation nonzero matrix elements are computed in less than one
minute.  We have previously discussed the problem
of finding the low-lying states of such a matrix\cite{Matrix}.


\begin{thebibliography}{1}
\bibitem{Kogut} J. B. Kogut and L. Susskind, {\em Phys. Rev.} {\bf D11}
                (1975) 395.
\bibitem{Gafix} J. B. Bronzan, {\em Phys. Rev.} {\bf D31} (1985) 2020.
\bibitem{Harm1} J. B. Bronzan and T. E. Vaughan, {\em Phys. Rev.} {\bf D44}
                (1991) 3264.
\bibitem{Matrix} J. B. Bronzan and T. E. Vaughan, {\em Phys. Rev.}
{\bf B39} (1989) 11724.
\end{thebibliography}
\end{document}